\documentclass{article}

\usepackage[preprint]{neurips_2025}

\usepackage[utf8]{inputenc} % allow utf-8 input
\usepackage[T1]{fontenc}    % use 8-bit T1 fonts
\usepackage{rotating}
\usepackage{url}            % simple URL typesetting
\usepackage{booktabs}       % professional-quality tables
\usepackage{amsfonts}       % blackboard math symbols
\usepackage{nicefrac}       % compact symbols for 1/2, etc.
\usepackage{microtype}      % microtypography
\usepackage[table,xcdraw]{xcolor}
\usepackage{tablefootnote}
\usepackage{graphicx}
\usepackage{wrapfig}
\usepackage{algorithm}
\usepackage{amsmath}
\usepackage{algpseudocode}
\usepackage{lscape}

\usepackage[colorlinks,linkcolor=black,citecolor=blue]{hyperref}

\title{SRDiffusion: Accelerate Video Diffusion Inference via Sketching-Rendering Cooperation}

\author{%
Shenggan Cheng$^{1}$ \quad Yuanxin Wei$^{2}$ \quad Lansong Diao$^{3}$ \quad Yong Liu$^1$ \quad Bujiao Chen$^{3}$ \\
\textbf{Lianghua Huang}$^{3}$ \quad \textbf{Yu Liu}$^{3}$ \quad \textbf{Wenyuan Yu}$^{3}$ \quad \textbf{Jiangsu Du}$^{2}$ \quad \textbf{Wei Lin}$^{3\dag}$ \quad \textbf{Yang You}$^{1\dag}$ \\
$^1$National University of Singapore \quad $^2$Sun Yat-sen University \quad $^3$Alibaba Group \\
$^{\dag}$ Corresponding authors
}

\begin{document}

\maketitle

\begin{abstract}
Leveraging the diffusion transformer (DiT) architecture, models like Sora, CogVideoX and Wan have achieved remarkable progress in text-to-video, image-to-video, and video editing tasks. Despite these advances, diffusion-based video generation remains computationally intensive, especially for high-resolution, long-duration videos. Prior work accelerates its inference by skipping computation, usually at the cost of severe quality degradation. In this paper, we propose SRDiffusion, a novel framework that leverages collaboration between large and small models to reduce inference cost. The large model handles high-noise steps to ensure semantic and motion fidelity (Sketching), while the smaller model refines visual details in low-noise steps (Rendering). Experimental results demonstrate that our method outperforms existing approaches, over 3$\times$ speedup for Wan with nearly no quality loss for VBench, and 2$\times$ speedup for CogVideoX. Our method is introduced as a new direction orthogonal to existing acceleration strategies, offering a practical solution for scalable video generation.
\end{abstract}

\section{Introduction}

With the rapid development of diffusion models, they have become the mainstream approach for generating high-quality images, audio, and video. Among them, DiT-based \cite{peebles2023scalable} video generation models have also advanced rapidly, including Sora \cite{brooks2024video}, CogVideoX \cite{yang2024cogvideox}, OpenSora \cite{zheng2024open}, Wan \cite{wang2025wan}, and others. These models have been widely applied in various tasks such as image-to-video generation, text-to-video generation, video editing \cite{wang2023videocomposer, jiang2025vace}, and video personalization \cite{wei2024dreamvideo}.

However, despite significant advancements in generation quality, diffusion-based video generation remains computationally expensive and time-consuming. The inference cost increases rapidly with model size, video resolution, and temporal duration. For instance, generating a 5-second 720p video using the Wan-14B model can take nearly an hour on a single NVIDIA A100 GPU. Prior acceleration works \cite{lv2024fastercache, zhao2024real, liu2024timestep} focus on the \textit{computation skipping} methods, caching certain diffusion steps or intermediate results to exploit similarities across different sampling stages.
Despite their limited speedup, they often lead to a noticeable decline in generation quality.

In this study, based on observations from the VBench evaluation of both large and small models, we find that the primary advantage of large models lies in their superior semantic capabilities, particularly in following instructions for composition and motion. However, the difference is much smaller in terms of detail quality (the "quality" dimension in VBench). On the other hand, small models have a significant advantage in runtime efficiency.

Building on these insights, we propose SRDiffusion, a novel approach to accelerate diffusion inference via sketching-rendering cooperation. Specifically, SRDiffusion will use large model during the high-noise steps to generate higher-quality structure and motion that better align with textual instructions (\textbf{\textit{Sketching}}), while use the small model during the low-noise steps to generate finer details (\textbf{\textit{Rendering}}), thereby accelerating the overall diffusion process. 
In addition, we design a metric to dynamically determine the switch from the sketching phase to the rendering phase, enabling a more flexible and guaranteed speed-quality trade-off.

The contributions of our paper are as follows:

\begin{itemize}

\item We reveal the distinct trade-offs in semantics, quality, and speed between large and small models, and highlight the potential for their cooperative use.

\item The introduction of sketching-rendering cooperation, a novel approach that leverages large models for sketching and small models for rendering, to accelerate video diffusion.

\item We design an adaptive switching metric to decide the time of switch from the sketching phase to rendering phase.

\item Experimental results demonstrate that our method outperforms existing approaches, over 3$\times$ speedup for Wan with nearly no quality loss for VBench, and 2$\times$ speedup for CogVideoX.

\end{itemize}
\section{Preliminaries and Related Works}

\subsection{Diffusion Process.}

Diffusion models \cite{ho2020denoising} simulate the gradual diffusion of data into Gaussian noise (the forward process) and the subsequent recovery of the original data from noise (the reverse process) to achieve the modeling and generation of complex data distributions. In the forward diffusion process, starting from a data point sampled from the real distribution, $x_0 \sim q(x)$, Gaussian noise $\epsilon_t$ is gradually added over $T$ steps to produce a sequence of increasingly noisy samples ${\{x_t\}_{t=1}^T}$, where the noise level is controlled by $\alpha_t$:

\begin{equation}
x_t = \sqrt{\alpha_t} x_{t-1} + \sqrt{1 - \alpha_t} \epsilon_t, \quad \epsilon_t \sim \mathcal{N}(0, I), \quad \alpha_t \in[0,1], \quad t=1, 2, ..., T
\label{eq:denoise}
\end{equation}

In the reverse diffusion process, a neural network is trained to approximate the conditional distribution $O(x_t,t)$, effectively learning how to denoise a sample at each step. 
The model iteratively removes noise, moving from $x_T$ back to a clean sample $x_0$, thereby generating new data consistent with the training distribution. A scheduler $\Phi$ determines how to exactly compute $x_{t-1}$ from $x_t$, $t$ and the output of the neural network $O(x_t,t)$: 

\begin{equation}
x_{t-1} = \Phi(x_t,t,O(x_t,t)), \quad t=T, ..., 2, 1
\label{eq:reverse}
\end{equation}

\subsection{Video Diffusion Transformer}

The video diffusion transformer consists of three primary components: a 3D Variational Autoencoder (3D VAE), a text encoder, and a diffusion transformer. The 3D VAE compresses the input video from the pixel space $V \in \mathbb{R}^{(1+T) \times H \times W \times 3}$ into a latent representation $x \in \mathbb{R}^{(1+T/C_T) \times H/C_H \times W/C_W}$, where $C_T, C_H, C_W$ denote the compression rates in the temporal, height, and width dimensions, respectively. This latent representation is then reshaped into a flattened sequence $z_{vision}$. The text encoder processes the input text into a corresponding latent embedding $z_{text}$. To embed text conditions, Wan uses cross-attention, whereas CogVideoX concatenates $z_{vision}$ and $z_{text}$ directly.

\subsection{Diffusion Inference Acceleration}

To accelerate diffusion inference, several studies have focused on designing more efficient schedulers, such as DDIM \cite{song2020denoising}. Others have explored advanced ODE or SDE solvers to improve sampling efficiency \cite{karras2022elucidating, lu2022dpm1, lu2022dpm}. In parallel, model distillation approaches aim to reduce inference time by training smaller models or models that require fewer sampling steps. For instance, \cite{wang2023videolcm, salimans2022progressive} employ distillation techniques to achieve high-quality generation with only a few steps. Additionally, some research efforts focus on improving model architecture \cite{xie2024sana, chen2024deep} or generative paradigm \cite{tian2024visual, gu2024dart, zhang2025packing, he2025neighboring} to enhance efficiency. However, these methods typically require fine-tuning or additional training, incurring extra computational costs.

Training-free methods can be broadly categorized into two strategies: skip-computation and system-level optimizations. Skip-computation techniques exploit redundancy across sampling steps, often leveraging caching mechanisms to accelerate inference \cite{lv2024fastercache, zhao2024real, liu2024timestep}. For example, T-GATE \cite{zhang2024cross} caches self- and cross-attention outputs at various stages, while PAB \cite{zhao2024real} selectively caches and broadcasts intermediate features based on attention block characteristics. TeaCache \cite{liu2024timestep} introduces timestep embedding-aware caching to bypass certain diffusion steps. On the other hand, system-level approaches, such as quantization \cite{zhang2024sageattention, deepgemm2025, li2024svdqunat} and parallelism \cite{fang2024usp, fang2024xdit, li2024distrifusion, zhao2024dsp}, aim to further reduce computational overhead through architectural and execution-level optimizations. 

This work proposes a new optimization direction that is orthogonal to the related studies mentioned above: leveraging collaboration between large and small models for diffusion inference. Specifically, the large model is used during the high-noise steps to generate higher-quality structure and motion that better align with textual instructions (\textbf{\textit{Sketching}}), while the small model is employed during the low-noise steps to generate finer details (\textbf{\textit{Rendering}}), thereby accelerating the overall process.

The cooperation concept is widely discussed in the context of LLM serving, known as speculative decoding \cite{leviathan2023fast, chen2023accelerating}. In contrast to auto-regressive models, where a smaller model is used for speculation and a larger one for verification, we propose the opposite for diffusion models: employing a larger model for sketching and a smaller model for rendering. The optimization principles are also quite different: speculative decoding improves the hardware utilization of large models through batch verification, whereas our method directly reduces computation by using a smaller model for certain steps. And our proposed approach aligns with certain edge-cloud system architectures, such as Hybrid-SLM-LLM \cite{hao2024hybrid} and HybridSD \cite{yan2024hybrid}, where a lightweight model on the edge collaborates with cloud-based models to reduce the overall generation cost. 
\section{Method}

\subsection{Motivation}

Based on evaluation results from VBench \cite{huang2024vbench} across both the semantic and quality dimensions, we observe that the Wan 14B model demonstrates a significant improvement in semantic dimension compared to the Wan 1.3B model. While the quality dimensions between the two models are relatively close, the larger model still maintains a slight advantage, as shown in Table \ref{tab:motivation-vbench}. However, there is a notable trade-off in inference latency: the Wan 1.3B model is over five times faster than the Wan 14B.

\begin{table}[bth]
\centering
\small
\caption{VBench scores and sample latency for 480p video on A800 for Wan.\tablefootnote{VBench scores are from \url{https://hf.co/spaces/Vchitect/VBench_Leaderboard} for Wan2.1 (2025-02-24) and Wan2.1-T2V-1.3B (2025-05-03). Latency is measured on a public cloud A800 instance.}}
\renewcommand\arraystretch{1.2}
\begin{tabular}{lcccc}
\toprule
         & Total Score & Quality Score & Semantic Score & Sample Latency   \\ \hline
Wan 14B  & 86.22       & 86.67         & 84.44          &       841\,s       \\
Wan 1.3B & 83.31       & 85.23         & 75.65          &       158\,s       \\ \bottomrule
\end{tabular}
\label{tab:motivation-vbench}
\end{table}

\begin{figure}[hbt]
\centering
\includegraphics[width=1.0\columnwidth]{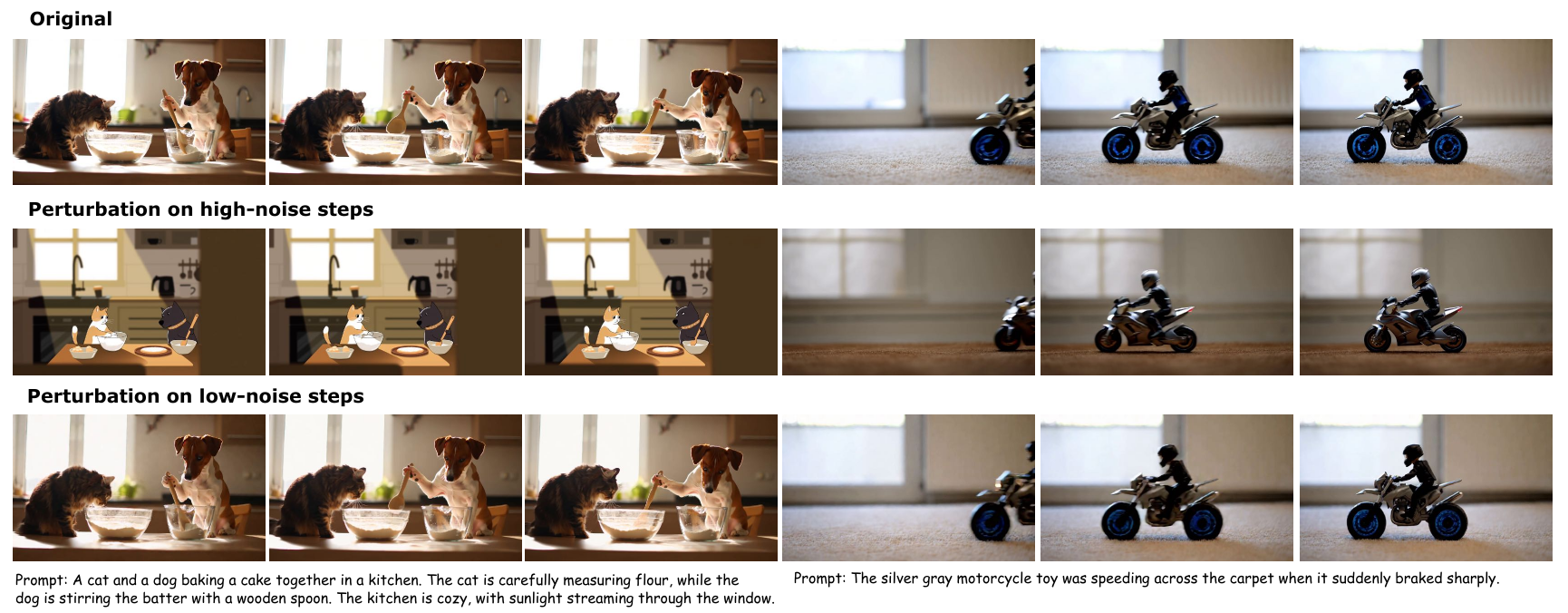}
\caption{Impact of perturbations at various diffusion steps on the quality of video frames.}
\label{fig:motivation-perturbation}
\end{figure}

To identify the most critical phase of the diffusion process for capturing semantics, we introduce perturbations in the form of biased Gaussian noise into the latent representation at different stages. As shown in Figure \ref{fig:motivation-perturbation}, perturbations introduced during the early high-noise steps (steps 0 to 10) lead to significant semantic changes, altering the overall structure and style of the video. In contrast, perturbations applied during the later low-noise steps (steps 10 to 50) result in only subtle variations in fine-grained details, such as the background chair in the first example or texture refinement in the second, while largely preserving the core semantics.

These observations motivate our two-stage approach. We propose using a larger model as a \textbf{\textit{Sketching Model}} to provide strong semantic guidance and ensure accurate content generation during the high-noise steps. Subsequently, a smaller model serves as a \textbf{\textit{Rendering Model}} to refine the output during the low-noise steps, accelerating the final generation. This hybrid strategy combines the strengths of both models, enabling high-quality results with reduced inference time.

\subsection{Sketching-Rendering Cooperation}

\begin{figure}[hbt]
\centering
\includegraphics[width=0.90\columnwidth]{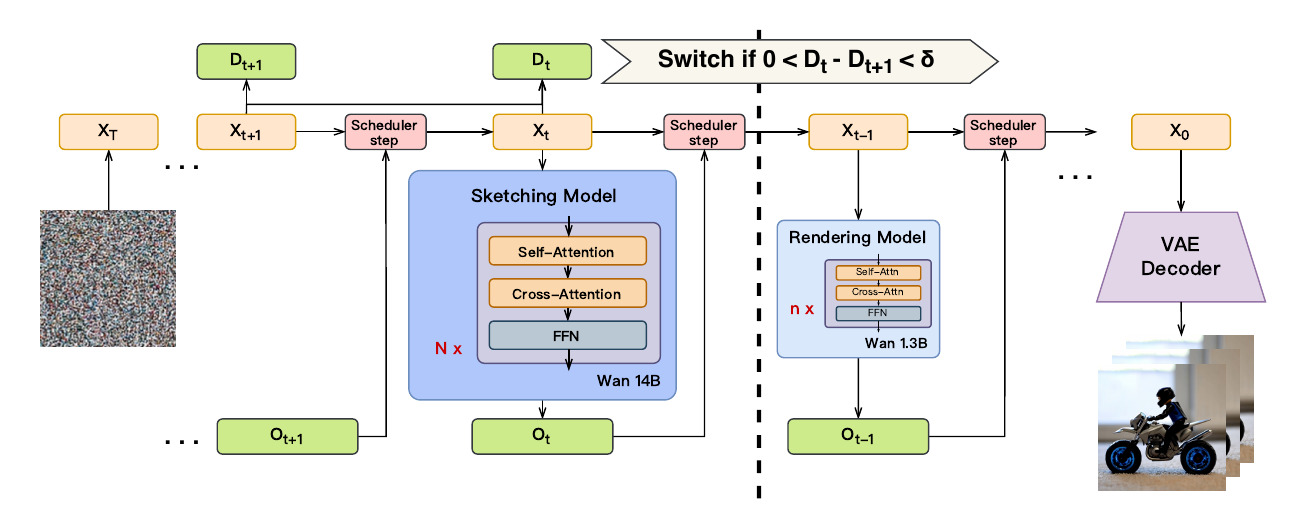}
\caption{Overview of Sketching-Rendering Cooperation. Taking pipeline of Wan model as an example, illustrates the pipeline switches from Wan 14B to Wan 1.3B at timestep $t$.}
\label{fig:overview}
\end{figure}

Our previous analysis highlighted that the early high-noise steps of the diffusion process are particularly important for semantic aspects such as composition and motion. During this phase, larger models demonstrate significantly stronger semantic capabilities compared to smaller ones. In contrast, the low-noise steps mainly focus on fine-grained details. Although smaller models are slightly less capable in this stage, they offer a clear advantage in terms of speed. Based on these observations, we propose the Sketching-Rendering Cooperation framework, which is illustrated in the overall architecture of Figure \ref{fig:overview}.

In this pipeline, the entire diffusion process generates a video based on the given input prompt. The prompt is first processed by a text encoder, and its encoded representation is used as a condition throughout every step of the diffusion process. The generation begins with a randomly initialized noise latent, which is initially handled by the sketching model. This model predicts the noise and updates the latent using a scheduler step. Then, an adaptive switching mechanism will determine whether to continue using the sketching model or switch to the rendering model. At a certain timestep $t$, once the mechanism decides that the rendering model can take over, the remaining diffusion steps are performed by it. At the end, the resulting latent is decoded into a video using a 3D VAE decoder.

Throughout the process, the collaboration between the sketching model and the rendering model ensures a balance between quality and efficiency. The sketching model preserves high-level semantics in the early phase, while the rendering model generates detailed content in the later phase with lower computational cost.

In most state-of-the-art models, such as Wan \cite{wang2025wan}, Hunyuan \cite{kong2024hunyuanvideo}, the 3D VAE is typically trained separately before training the DiT. Within the same model family, different model sizes generally share the same VAE, which uses identical compression parameters and maintains a consistent latent space. As a result, when switching from the sketching model to the rendering model, the latent tensor shapes remain the same, and no additional alignment is required. The corresponding pseudocode is provided in Algorithm \ref{algo:sr-diff-algo}. The pseudocode shows the sketching-rendering cooperation for Wan with classifier-free guidance.

\begin{algorithm}[H]
\small
\caption{Diffusion Inference Process for Sketching-Rendering Cooperation for Wan.}
\begin{algorithmic}[1]
\State Initialize latent variable $\mathbf{z}$
\State Set initial model: $\text{ExecModel} \gets \text{SketchingModel}$
\For{each timestep $t$ in $\{T, \dots, 1\}$}
    \State Predict conditional noise: $\hat{\epsilon}_{\text{cond}} \gets \text{ExecModel}(\mathbf{z}, t, \text{condition})$
    \State Predict unconditional noise: $\hat{\epsilon}_{\text{uncond}} \gets \text{ExecModel}(\mathbf{z}, t, \text{no condition})$
    \State Apply guidance: $\hat{\epsilon} \gets \hat{\epsilon}_{\text{uncond}} + s \cdot (\hat{\epsilon}_{\text{cond}} - \hat{\epsilon}_{\text{uncond}})$
    \State Update latents: $\mathbf{z} \gets \text{SchedulerStep}(\hat{\epsilon}, t, \mathbf{z})$
    \If{ExecModel is SketchingModel and switch\_condition($\mathbf{z}$)} \Comment{Switch Condition see Sec. \ref{sec:switch}}
        \State Switch to rendering model: $\text{ExecModel} \gets \text{RenderingModel}$
    \EndIf
\EndFor
\State \Return Final generated sample $\mathbf{z}_0$
\end{algorithmic}
\label{algo:sr-diff-algo}
\end{algorithm}

\subsection{Adaptive Switch}
\label{sec:switch}

% We find that for different prompts, to ensure the similarity of the output of original sketching model, we need dynamically decide in which step to switch from sketching model to rendering model.
% Following \cite{liu2024timestep} and \cite{cache_me}, we observe the minor change of the model output across steps in the late denoising stage.
% We apply the relative L1 distance to characterize the difference of the model output across steps. 
% For instance, the output difference between timestep $t$ and the timestep $t-1$  is calculated as follows:
% \begin{equation}
% D_t = tanh(\frac{||O_t-O_{t-1}||_1}{||O_{t-1}||_1}) 
% \label{eq:diff}
% \end{equation}
% where $O_t$ indicates the model output at timestep $t$.
% We apply $tanh$ to scaling the absolute value to the range of (0,1).
% Based on Equation \ref{eq:diff}, we visualize the output changes in Wanx 14B and CogVideo 5B in Figure \ref{fig:diff_values}(a)(c). 

\begin{figure}[hbt]
\centering
\includegraphics[width=1\columnwidth]{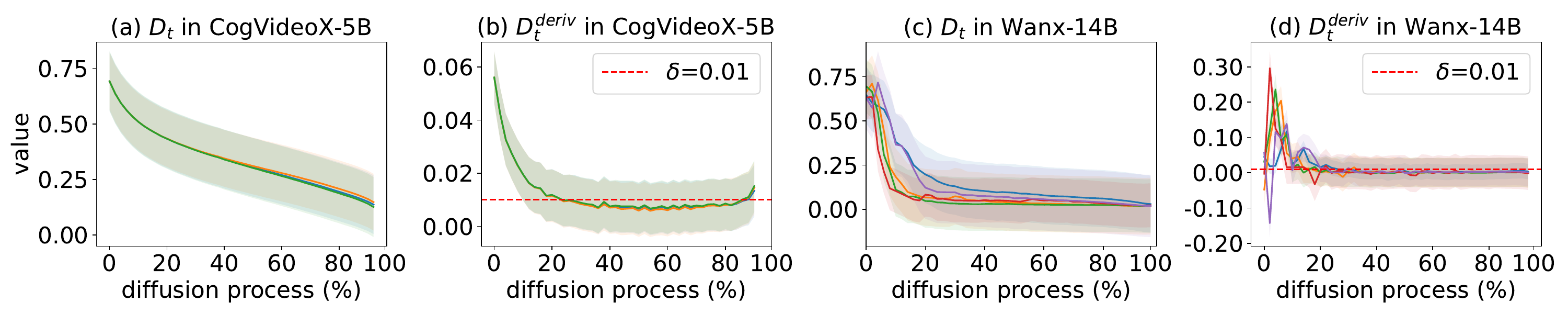}
\caption{Predicted noise difference across denoising steps in Wan-14B 480p and CogVideoX-5B 480p. Different colors represent the value of different prompts.}
\label{fig:diff_values}
\end{figure}

% \djs{
To ensure a consistent effect for different prompts, we dynamically determine the optimal switching step from the sketching model to the rendering model.
Following \cite{liu2024timestep} and \cite{cache_me}, we observe that the predicted Gaussian noise change diminishes during the diffusion process, and the second-order derivative of this change continuously decreases and gradually stabilizes.
We use the relative L1 distance to characterize the difference of the denoised sample across  steps as follows:
\begin{equation}
D_t = tanh(\frac{||x_t-x_{t+1}||_1}{||x_{t+1}||_1}) 
\label{eq:diff}
\end{equation}
where $x_t$ indicates the denoised sample at timestep $t$ following \ref{eq:denoise} and $tanh$ is applied to scale the absolute value to the range of (0,1).
The denoised sample changes $D_t$ of multiple prompts throughout the diffusion process in Wanx 14B and CogVideoX 5B are illustrated in Figure \ref{fig:diff_values}(a)(c), and the their second-order derivative are illustrated in Figure \ref{fig:diff_values}(b)(d), respectively. 

% }
% 应该不准确，看起来二阶导是不断下降然后平稳，yuanxin看看。

% We observe that, the absolute value of $D_t$ varies greatly for different prompts in the same model, but the changing trend across steps keeps the same. In order to better characterize the output changes for all prompts, we further compare the output difference with the previous step:
% \begin{equation}
% D^{self}_t = D_t - D_{t-1} < \delta
% \label{eq:self_diff}
% \end{equation}
% As illustrated in Figure \ref{fig:diff_values}(b)(d), the absolute value of $D^{self}_t$ across different prompts are similar.
% Finally, we use $D^{self}_t$ as the metric to decide the timestep of switching from sketching model to rendering model.
% During runtime, we record $D^{self}_t$ of each timestep in the early denoising stage and compare it with the threshold $\delta$. Once Equation \ref{eq:self_diff} is satisfied, we switch to the rendering model.

During runtime, we record the second-order derivative of predicted noise $D^{deriv}_t$ of each timestep and compare it with the threshold $\delta$. 
Besides, to ensure the video quality, we set \texttt{FIX\_STEP} as the minimum scheduler steps to execute the sketching model, which is set to 5 in our experiments.
Note that as the reverse diffusion process progresses, the index of denoising timestep $t$ decreases, while that of the scheduler step $\tau$ increases.
Once the following three conditions are all satisfied, we switch to the rendering model.
\begin{align}
    &D^{deriv}_t = D_t - D_{t+1} < \delta \\
    &D^{deriv}_t > 0  \\
    &\tau \geq \texttt{FIX\_STEP} 
\end{align}

In practice, we provide the distribution of steps selected by the adaptive switching strategy under different values of $\delta$ in the Appendix \ref{sec:distribution-adaptive}.
\section{Experiments}

\subsection{Experimental Setting}
\label{sec:exp_setting}

\textbf{Models.} We conduct experiments on multiple video generation models, including Wan and CogVideoX, which provide two model sizes: Wan includes 14B and 1.3B variants, while CogVideoX includes 5B and 2B variants. As model families, they share the same VAE within each family.

\textbf{Baselines.} For baseline methods, we select PAB \cite{zhao2024real} and TeaCache \cite{liu2024timestep}, both of which are specifically designed to accelerate video diffusion models through caching mechanisms. These techniques are conceptually similar to our approach in that they aim to skip redundant computations, whereas our method switches to a smaller model for computation reduction.

\textbf{Metrics.} For quality evaluation, we utilize VBench \cite{huang2024vbench} to evaluate the generation quality. We use VBench standard prompt set and generate 5 videos with different seeds for each prompt. In addition, we report standard perceptual and pixel-level similarity metrics, including Learned Perceptual Image Patch Similarity (LPIPS) \cite{zhang2018unreasonable}, Structural Similarity Index Measure (SSIM) \cite{wang2004image}, and Peak Signal-to-Noise Ratio (PSNR). For efficiency evaluation, we measure the inference latency per sample as the key performance indicator.

\textbf{Experiment Details.} All experiments are conducted on public cloud instances with NVIDIA A800 80GB GPUs using PyTorch with bfloat16 mixed-precision. We enable FlashAttention \cite{dao2022flashattention} by default to accelerate attention computation.

\subsection{Main Results}

\textbf{Quantitative Comparison.} Table \ref{tab:main-results} presents a quantitative evaluation of video generation quality, similarity, and inference speed using VBench. Prompt extension is performed using Qwen2.5 following the instructions from Wan. We adopt two variants to explore different quality-speed trade-offs, controlled by $\delta$, the smaller $\delta$ will switch to rendering model later and get more fidelity results from original sketching model. 

For TeaCache, we use the open-source implementation and adjust the \texttt{l1\_distance\_thresh} parameter to balance quality and speed. Additionally, we adopt the PAB implementation from HuggingFace Diffusers \cite{von-platen-etal-2022-diffusers}, tuning both \texttt{block\_skip\_range} and \texttt{timestep\_skip\_range} to manage the quality-speed trade-off. The baseline models are tuned to ensure they fall within a comparable quality-speed spectrum. Further experimental results, and VBench scores across individual dimensions are provided in Appendix \ref{sec:detail-vbench-score}.

\begin{table}[hbt]
\centering
\small
\caption{Quality results for video generation quality, similarity and inference speed. Similarity metrics are calculated against the original larger model (Wan 14B and CogVideoX 5B) results.}
\renewcommand\arraystretch{1.1}
\begin{tabular}{rcccccccc}
\toprule
\multicolumn{1}{c}{Method} & \multicolumn{3}{c}{\begin{tabular}[c]{@{}c@{}}VBench↑\\ (Total, Quality, Sematic)\end{tabular}} & \multicolumn{3}{c}{\begin{tabular}[c]{@{}c@{}}Similarity\\ (LPIPS↓, PSNR↑, SSIM↑)\end{tabular}} & Latency & Speedup \\ \hline
\multicolumn{9}{c}{Wan (832$\times$480, 5s, $T=50$)}   \\ \hline
\rowcolor{gray!10} Wan 14B                 & 84.05 & 85.04 & 80.09 & - & - & - & 841\,s & 1$\times$ \\
\rowcolor{gray!10} Wan 1.3B                & 83.12 & 84.28 & 78.50 & - & - & - & 158\,s & -  \\
PAB$_{8,[100-970]}$        & 81.83 & 83.14 & 76.59 & 0.247 & 19.94 & 0.706 & 506\,s & 1.66$\times$ \\
TeaCache$_{0.14}$          & 83.95 & 84.85 & 80.34 & 0.244 & 18.60 & 0.688 & 579\,s & 1.45$\times$ \\
TeaCache$_{0.2}$           & 83.69 & 84.76 & 79.42 & 0.331 & 16.59 & 0.620 & 427\,s & 1.96$\times$ \\
\rowcolor{green!5} \textbf{SRDiff\,($\delta=0.01$)}   & 84.06 & 84.97 & \textbf{80.40} & \textbf{0.197} & \textbf{20.53} & \textbf{0.734} & 296\,s & 2.84$\times$ \\
\rowcolor{green!5} \textbf{SRDiff\,($\delta=0.03$)}   & \textbf{84.12} & \textbf{85.08} & 80.29 & 0.233 & 19.32 & 0.700 & \textbf{262\,s} & \textbf{3.21$\times$} \\ \hline
\multicolumn{9}{c}{CogVideoX (720$\times$480, 6s, $T=50$)}                                                                                                                    \\ \hline
\rowcolor{gray!10} CogVideoX 5B & 80.89 & 82.20 & 75.61 & - & - & - & 213\,s & 1$\times$ \\
\rowcolor{gray!10} CogVideoX 2B & 80.04 & 81.39 & 74.65 & - & - & - & 75\,s  & - \\
PAB$_{8,[100-900]}$                & 76.84 & 78.98 & 68.27 & 0.370 & 17.68 & 0.670 & 105\,s  &  2.03$\times$  \\
TeaCache$_{0.1}$       & 80.15 & 81.22 & \textbf{75.87} & 0.239 & 20.42 & 0.741 & 139\,s & 1.53$\times$ \\
TeaCache$_{0.15}$      & 79.16 & 80.43 & 74.09 & 0.348 & 17.64 & 0.662 & 106\,s & 2.01$\times$ \\
\rowcolor{green!5} \textbf{SRDiff\,($\delta=0.01$)}   & \textbf{80.85}  & \textbf{82.22} & 75.38 & \textbf{0.177} & \textbf{22.93} & \textbf{0.795} & 117\,s & 1.82$\times$ \\
\rowcolor{green!5} \textbf{SRDiff\,($\delta=0.015$)}  & 80.51 & 81.91 & 74.91 & 0.260 & 19.77 & 0.710 & \textbf{104\,s} & \textbf{2.05$\times$} \\ \bottomrule
\end{tabular}
\label{tab:main-results}
\end{table}

For the Wan-based models, SRDiffusion (denoted as SRDiff in the table)  achieves significant speedup and consistently outperforms all baselines. Even under the speed-oriented configuration ($\delta=0.03$), SRDiffusion achieves higher VBench scores than all baselines, even slightly surpassing the original large model Wan 14B, while reducing latency by more than 3×. In terms of similarity metrics, the quality-oriented variant ($\delta=0.01$) delivers better overall performance.  Overall, SRDiffusion demonstrates a clear advantage in acceleration for Wan-based models, with no observable degradation in VBench quality scores.

In the CogVideoX setting, SRDiffusion also demonstrates competitive performance. At $\delta=0.01$, it nearly matches the original model in VBench score (80.85 vs. 80.89) while achieving the best similarity metrics across all evaluated methods, with a 1.82× speedup. The $\delta=0.015$ variant offers a slight reduction in quality (VBench 80.51) but achieves a higher speedup of 2.05×, outperforming other baselines with comparable runtime, including PAB and TeaCache-0.15.

The relatively lower acceleration ratio observed for CogVideoX, compared to Wan, is primarily due to the smaller performance gap between the large and small CogVideoX models. Additionally, the switch step in CogVideoX occurs later during inference, which further limits speed gains.

\begin{figure}[hbt]
\centering
\includegraphics[width=0.95\columnwidth]{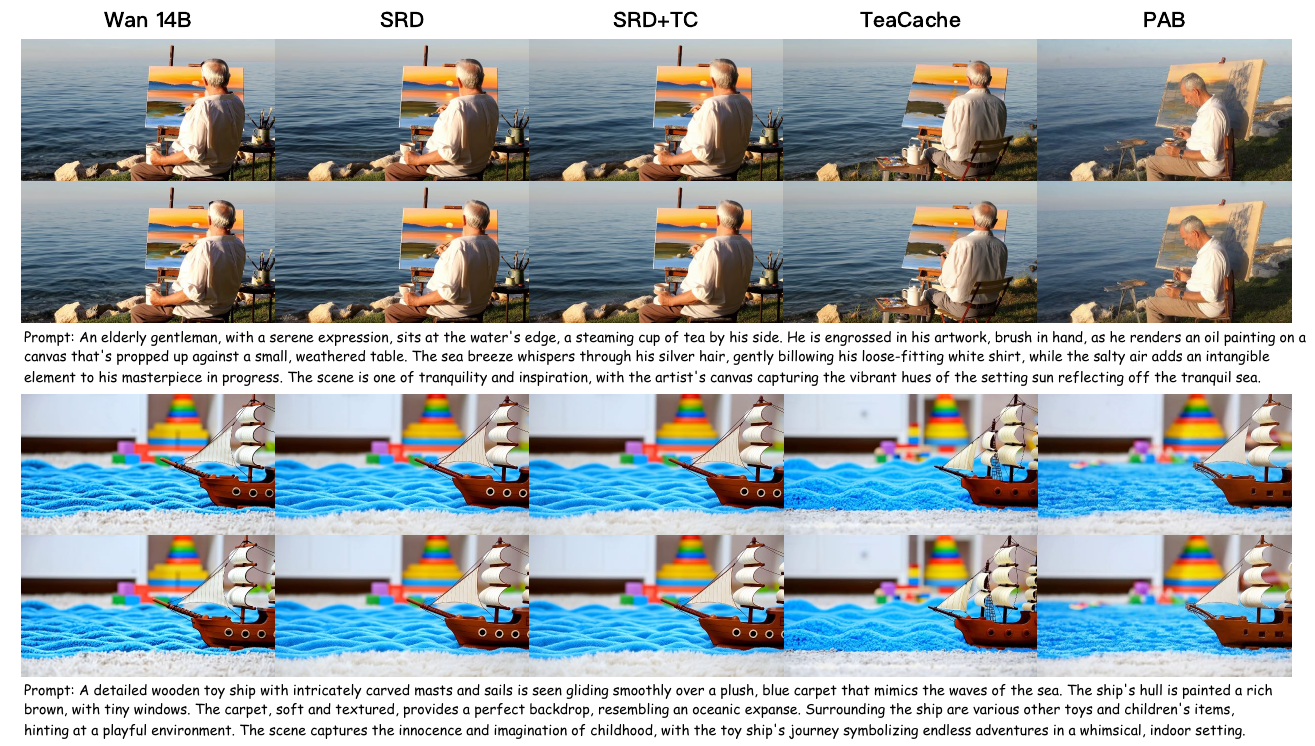}
\caption{Visualization Results. We compare the generation quality between original model, our method and baselines. (SDR: SRDiffusion, TC: TeaCache)}
\label{fig:main-viz-results}
\end{figure}

\textbf{Visualization Results.} As shown in Figure \ref{fig:main-viz-results}, we present video results generated by the baseline, our method, and the original Wan model for comparison, using selected challenging prompts \footnote{The prompts used for Figure \ref{fig:main-viz-results} are from \url{https://github.com/THUDM/CogVideo/blob/main/resources/galary_prompt.md}}. The visualizations demonstrate that our method more faithfully preserves the composition of the original model and achieves superior detail generation quality. The "SRD+TC" configuration in the figure illustrates the complementary use of our method with TeaCache, which will be discussed further in Section \ref{sec:plug-and-play}. Additional visualization results for Wan and CogVideoX on VBench prompts and more challenging prompts are provided in Appendix \ref{sec:appendix-viz}.

\subsection{Complementary Use with Other Optimizations}
\label{sec:plug-and-play}

SRDiffusion operates as a plug-and-play solution and can be seamlessly integrated with various optimization techniques, such as caching mechanisms or system-level enhancements. In our implementation, we integrate TeaCache with SRDiffusion to further improve efficiency. To ensure stability during the sketching stage, TeaCache is activated only in the rendering stage. We evaluate the combined method, SRDiffusion+TeaCache, on the Wan and CogVideoX using VBench, with the results summarized in Table~\ref{tab:combine}.

As shown, SRDiffusion+TeaCache consistently outperforms both baselines in terms of latency and speedup, while maintaining competitive quality. On the Wan model, it achieves a 3.91× speedup with the lowest latency (215s), while preserving high semantic and visual fidelity (VBench: 83.82, LPIPS: 0.194, SSIM: 0.741). Similarly, on CogVideoX, SRDiffusion+TeaCache offers the fastest runtime (107s) and the highest speedup (1.99×), with negligible quality trade-offs compared to SRDiffusion alone. These results demonstrate the effectiveness and efficiency of SRDiffusion when combined with other cache-based optimizations. Compared to TeaCache, the main advantage of our method is that we don't skip any steps during the sketching stage, thereby achieving stronger semantic alignment.

\begin{table}[hbt]
\centering
\small
\caption{VBench and Similariry Results for SRDiffusion+TeaCache.}
\renewcommand\arraystretch{1.1}
\begin{tabular}{rcccccccc}
\toprule
\multicolumn{1}{c}{Method} & \multicolumn{3}{c}{\begin{tabular}[c]{@{}c@{}}VBench↑\\ (Total, Quality, Sematic)\end{tabular}} & \multicolumn{3}{c}{\begin{tabular}[c]{@{}c@{}}Similarity\\ (LPIPS↓, PSNR↑, SSIM↑)\end{tabular}} & Latency & Speedup \\ \hline
\multicolumn{9}{c}{Wan (832$\times$480, 5s, $T=50$)}   \\ \hline
TeaCache$_{0.14}$          & 83.95 & 84.85 & 80.34 & 0.244 & 18.60 & 0.688 & 579\,s & 1.45$\times$ \\
SRDiff\,($\delta=0.01$)   & 84.06 & 84.97 & 80.40 & 0.197 & 20.53 & 0.734 & 296\,s & 2.84$\times$ \\
\rowcolor{green!5} \textbf{SRDiff+TeaCache}  & 83.82 & 84.80 & 79.87 & 0.194 & 20.88 & 0.741 & \textbf{215\,s} & \textbf{3.91$\times$} \\ \hline
\multicolumn{9}{c}{CogVideoX (720$\times$480, 6s, $T=50$)}                                                                                                                   \\ \hline
TeaCache$_{0.1}$       & 80.15 & 81.22 & 75.87 & 0.239 & 20.42 & 0.741 & 139\,s & 1.53$\times$ \\
SRDiff\,($\delta=0.01$)   & 80.85  & 82.22 & 75.38 & 0.177 & 22.93 & 0.795 & 117\,s & 1.82$\times$ \\
\rowcolor{green!5} \textbf{SRDiff+TeaCache}  & 80.24 & 81.48 & 75.28 & 0.187 & 22.56 & 0.791 & \textbf{107\,s} & \textbf{1.99$\times$} \\
\bottomrule
\end{tabular}
\label{tab:combine}
\end{table}

\begin{figure}[hbt]
\centering
\includegraphics[width=0.98\columnwidth]{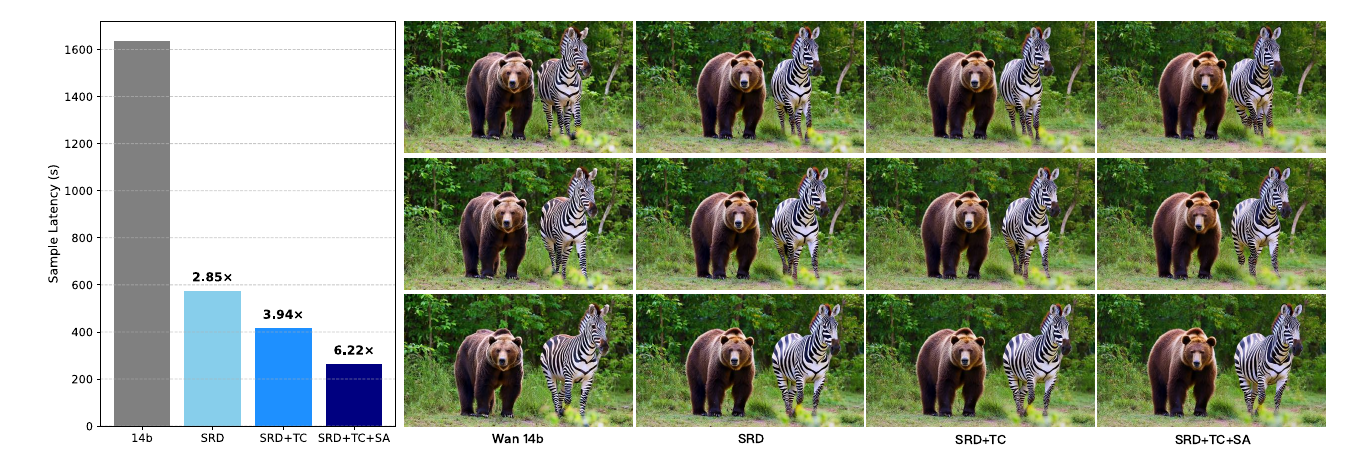}
\caption{SRDiffusion combined with TeaCache and SageAttention achieves over 6× speedup on a single NVIDIA H20. (SDR: SRDiffusion, TC: TeaCache, SA: SageAttention)}
\label{fig:combine}
\end{figure}

To further accelerate the process, we incorporate FP8 Attention by adopting SageAttention \cite{zhang2024sageattention}. The resulting speedup and a sample video (from VBench) are presented in Figure~\ref{fig:combine}. This experiment was conducted on the NVIDIA H20 platform, as SageAttention delivers notable performance improvements only on architectures newer than Hopper. As illustrated, the combination achieves a 6.22× speedup while maintaining consistent semantics and comparable visual quality.

\subsection{Analysis of Adaptive Switch}

\begin{figure}[hbt]
\centering
\includegraphics[width=0.98\columnwidth]{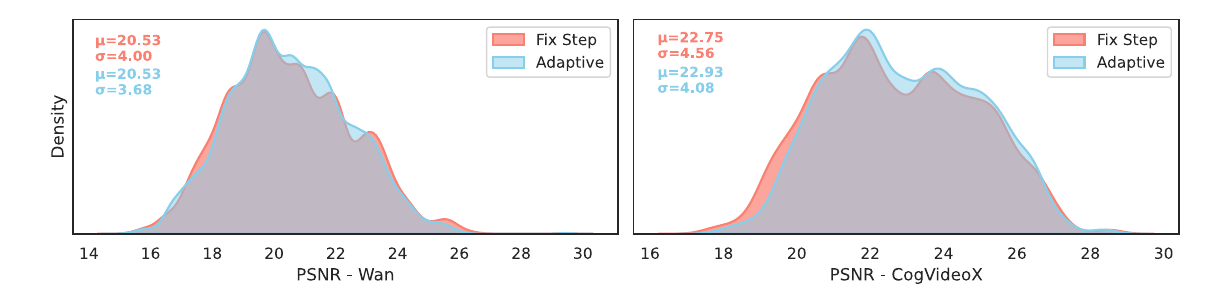}
\caption{PSNR Distribution for Adaptive Swtich and Fixed-Step Switch.}
\label{fig:adapt-fix-psnr}
\end{figure}

To illustrate the effectiveness of the adaptive switch mechanism, we compare it against a fixed-step switching baseline. Under the $\delta = 0.01$ setting, the average switching step is around 10 for Wan and 15 for CogVideoX. We therefore set these as fixed switching points in the baseline to ensure equivalent acceleration. The corresponding quality and similarity metrics are reported in Table \ref{tab:adaptive-fix}, and the distribution of PSNR values is shown in Figure \ref{fig:adapt-fix-psnr}.

Compared to the Fixed-Step Switch, the adaptive switch achieves nearly identical VBench scores but offers a slight advantage in similarity metrics, with the improvement being more pronounced on CogVideoX. By examining the distribution of PSNR values, we observe that adaptive switch results in lower variance, indicating more consistent similarity scores. In challenging cases where similarity is harder to maintain, the adaptive mechanism tends to delay the switch, thereby improving generation quality.

\begin{table}[hbt]
\centering
\small
\caption{Comparison of Adaptive Switch and Fixed-Step Switch.}
\renewcommand\arraystretch{1.1}
\begin{tabular}{rcccccc}
\toprule
\multicolumn{1}{c}{Method} & \multicolumn{3}{c}{\begin{tabular}[c]{@{}c@{}}VBench↑\\ (Total, Quality, Sematic)\end{tabular}} & \multicolumn{3}{c}{\begin{tabular}[c]{@{}c@{}}Similarity\\ (LPIPS↓, PSNR↑, SSIM↑)\end{tabular}} \\ \hline
\multicolumn{7}{c}{Wan (832$\times$480, 5s, $T=50$)}   \\ \hline
Fix Step\,($T=10$) & 84.05 & 84.96 & 80.40 & 0.196 & 20.53 & 0.732  \\ 
\rowcolor{green!5} Adaptive Switch($\delta=0.01$)   & 84.06 & 84.97 & 80.40 & 0.197 & 20.53 & 0.734  \\ \hline
\multicolumn{7}{c}{CogVideoX (720$\times$480, 6s, $T=50$)}                                                                                                                   \\ \hline
Fix Step\,($T=15$)  & 80.85 & 82.22 & 75.39 & 0.182 & 22.75 & 0.789 \\ 
\rowcolor{green!5} Adaptive Switch\,($\delta=0.01$)   & 80.85  & 82.22 & 75.38 & 0.177 & 22.93 & 0.795 \\ 
\bottomrule
\end{tabular}
\label{tab:adaptive-fix}
\end{table}

\subsection{Scaling to Higher Resolution}

Table \ref{tab:wan-720p} presents the VBench and similarity evaluation results for the Wan model at 720p resolution. Since CogVideoX only supports a maximum resolution of 480p, we don't include CogVideoX for evaluation. As shown in the table, SRDiffusion achieves the highest VBench scores across all submetrics and significantly outperforms the baselines in similarity metrics, indicating both higher perceptual and pixel-level fidelity. Moreover, it offers a 2.84× speedup over Wan 14B with much lower latency, demonstrating strong efficiency–quality tradeoffs.

\begin{table}[hbt]
\centering
\small
\caption{VBench and Similariry Results for Wan 720p.}
\renewcommand\arraystretch{1.1}
\begin{tabular}{rcccccccc}
\toprule
\multicolumn{1}{c}{Method} & \multicolumn{3}{c}{\begin{tabular}[c]{@{}c@{}}VBench↑\\ (Total, Quality, Sematic)\end{tabular}} & \multicolumn{3}{c}{\begin{tabular}[c]{@{}c@{}}Similarity\\ (LPIPS↓, PSNR↑, SSIM↑)\end{tabular}} & Latency & Speedup \\ \hline
\multicolumn{9}{c}{Wan (1280$\times$720, 5s, $T=50$)}   \\ \hline
\rowcolor{gray!10}Wan 14B   & 83.66 & 84.66 & 79.69 & - & - & - & 3130\,s & - \\
\rowcolor{gray!10}Wan 1.3B   & 83.57 & 84.69 & 79.06 & - & - & - & 637\,s & - \\
TeaCache$_{0.2}$ & 83.45 & 84.51 & 79.22 & 0.331 & 17.04 & 0.668 & 1572\,s & 1.99$\times$ \\
\rowcolor{green!5} \textbf{SRDiff\,($\delta=0.01$)} & \textbf{83.72} & \textbf{84.71} & \textbf{79.80} & \textbf{0.216} & \textbf{21.43} & \textbf{0.767} & \textbf{1104\,s} & \textbf{2.84$\times$} \\
\bottomrule
\end{tabular}
\label{tab:wan-720p}
\end{table}
\section{Discussion and Conclusions}

In conclusion, SRDiffusion offers a practical and effective solution to the computational challenges of diffusion-based video generation. By leveraging the semantic strengths of large models during the early, high-noise stages and the efficiency of small models during the later, low-noise stages, SRDiffusion significantly reduces inference time while preserving generation quality. SRDiffusion achieved more than 3$\times$ acceleration on Wan without any loss in VBench quality. It also achieved over 2$\times$ acceleration on CogVideoX. Additionally, SRDiffusion can be used alongside other methods to achieve even higher acceleration.

Currently, this work only focuses on sketching-rendering cooperation within the same family of models using a shared VAE, and thus cannot be generalized to arbitrary combinations of different models. In future work, we plan to explore more flexible architectures that enable cross-model cooperation by aligning latent spaces, thereby improving compatibility and scalability across diverse model types.

{
\small
\bibliographystyle{plain}
\bibliography{references}
}

%%%%%%%%%%%%%%%%%%%%%%%%%%%%%%%%%%%%%%%%%%%%%%%%%%%%%%%%%%%%

\newpage
\appendix

\section{Technical Appendices and Supplementary Material}
% Technical appendices with additional results, figures, graphs and proofs may be submitted with the paper submission before the full submission deadline (see above), or as a separate PDF in the ZIP file below before the supplementary material deadline. There is no page limit for the technical appendices.

\subsection{Further Experimental Results and VBench Score for Each Dimension}
\label{sec:detail-vbench-score}

Tables \ref{tab:detail-wan-vbench} and \ref{tab:detail-cogvideo-vbench} present the detailed VBench scores of Wan and CogVideoX across various dimensions. These tables also report results for a broader range of parameter configurations than those shown in the Main Results Table \ref{tab:main-results} in the main text. It can be observed that SRDiffusion demonstrates a clear speed advantage over all baseline configurations. Moreover, its VBench scores are closer to those of larger models, and on Wan model, it even slightly surpasses the 14B model in certain configurations.

\begin{table}[hbt]
\centering
\scriptsize
\caption{VBench Score for all dimensions, Wan Model.}
\renewcommand\arraystretch{1.1}
\setlength{\tabcolsep}{2pt}
\begin{tabular}{rccccccccccc}
\toprule
VBench Scores   & 14b   & 1.3b  & PAB$_{C1}$ & PAB$_{C2}$  & PAB$_{C3}$ & TC$_{C1}$ & TC$_{C2}$ & SRD$_{C1}$ & SRD$_{C2}$ & SRD$_{C3}$ & SRD$_{TC}$ \\ \hline
\textbf{total score}            & 84.05 & 83.12 & 83.13 & 82.42   &    81.83     & 83.95          & 83.69      & 83.87  & 84.06        & 84.12        & 83.82                                                                 \\ 
\textbf{speedup}            & 1$\times$ & - & 1.04$\times$  & 1.66$\times$ & 1.81$\times$ &  1.45$\times$      &   1.96$\times$ & 2.23$\times$ &  2.84$\times$  &  3.21$\times$   &  3.91$\times$              \\ \hline
\rowcolor{green!10} \textbf{semantic score}         & 80.09 & 78.5  & 79.70 & 78.16    &   76.59     & 80.34          & 79.42    &  79.85   & 80.4         & 80.29        & 79.87                                                                 \\
object class           & 93.09 & 90.87 & 92.52 & 90.47    &   88.58     & 93.01          & 91.04    &  92.47   & 92.44        & 92.1         & 92.5                                                                  \\
multiple objects       & 81.16 & 77.26 & 79.79 & 77.04    &   69.59     & 80.79          & 78.32    &  80.37   & 80.93        & 81.08        & 81.2                                                                  \\
human action           & 98    & 95.2  & 97.80 & 97.6     &  96.40      & 97.4           & 97       &  97.80   & 97.8         & 97.8         & 97.2                                                                  \\
color                  & 84.6  & 86.25 & 84.43 & 83.67    &  83.61      & 83.63          & 85.88    &  84.03   & 85.82        & 85.1         & 84.8                                                                  \\
spatial relationship   & 78.65 & 76.66 & 76.53 & 74.42    &  72.56      & 80.32          & 77.28    &   78.73  & 79.66        & 79.53        & 78.86                                                                 \\
scene                  & 66.58 & 66.97 & 67.72 & 65.06    &  65.17      & 69.81          & 66.65    &  66.42   & 68.23        & 68.18        & 66.61                                                                 \\
appearance style       & 77.2  & 73.3  & 77.02 & 76.18    &  76.39      & 76.81          & 77.34   &   76.74   & 76.49        & 76.35        & 76                                                                    \\
temporal style         & 68.35 & 67.75 & 68.35 & 66.9     &  65.80      & 68.21          & 67.83   &   68.71   & 68.76        & 68.87        & 68.46                                                                 \\
overall consist    & 73.16 & 72.25 & 73.13 & 72.12     &   71.18    & 73.05          & 73.41     &  73.35  & 73.43        & 73.57        & 73.19     \\
\rowcolor{orange!10} \textbf{quality score}          & 85.04 & 84.28 & 83.98 & 83.49    &   83.14    & 84.85          & 84.76     &  84.87  & 84.97        & 85.08        & 84.8                                                                  \\
subject consist    & 93.91 & 93.44 & 93.79 & 93.53     &   93.22    & 93.84          & 93.75     &  93.94  & 94           & 93.94        & 93.97                                                                 \\
background consist & 97.06 & 96.55 & 95.00 & 94.34    &   94.14     & 97.08          & 97.09    &  96.99   & 97.01        & 97.06        & 96.98                                                                 \\
temporal flickering    & 96.84 & 97.63 & 97.14 & 97.11    &    97.17    & 96.82          & 97.03    &  96.79   & 96.79        & 96.79        & 96.68                                                                 \\
motion smoothness      & 93.76 & 92.9  & 94.10 & 94.72    &   94.82     & 93.86          & 93.89    &   93.69  & 93.76        & 93.79        & 94                                                                    \\
dynamic degree         & 37.5  & 37.36 & 35.97 & 35.28   &    35.00     & 36.95          & 37.09    &   37.22  & 37.78        & 38.61        & 37.22                                                                 \\
aesthetic quality      & 66.9  & 64.35 & 63.16 & 62.05   &    61.34    & 66.75          & 66.54     &  66.28  & 66.22        & 66.17        & 66.08                                                                 \\
imaging quality        & 66.76 & 65.57 & 66.72 & 65.64   &   64.72      & 66.22          & 65.53    &  66.76   & 66.77        & 66.64        & 66.3                                                                  \\
\bottomrule                 
\end{tabular}
\label{tab:detail-wan-vbench}
\end{table}

\begin{table}[hbt]
\centering
\scriptsize
\caption{VBench Score for all dimensions, CogVideoX Model.}
\renewcommand\arraystretch{1.1}
\setlength{\tabcolsep}{2pt}
\begin{tabular}{rcccccccccccc}
\toprule
VBench Scores   & 5b    & 2b  & PAB$_{C1}$  & PAB$_{C2}$ & TC$_{C1}$ & TC$_{C2}$ & TC$_{C3}$ & SRD$_{C1}$ & SRD$_{C2}$ & SRD$_{C3}$ & SRD$_{C4}$ & SRD$_{TC}$ \\ \hline
\textbf{total score}            & 80.89 & 80.04 & 80.68 & 76.84            & 80.15        & 79.16         & 78.15      & 80.83 & 80.85        & 80.51         & 80.12        & 80.24                       \\
\textbf{speedup}            & 1$\times$ & - & 1.36$\times$ & 2.03$\times$ & 1.53$\times$  & 2.01$\times$ & 2.39$\times$ & 1.65$\times$ & 1.82$\times$  & 2.05$\times$  &  2.32$\times$   &  1.99$\times$                 \\ \hline
\rowcolor{green!10} \textbf{semantic score}         & 75.61 & 74.65 & 75.13 & 68.27            & 75.87        & 74.09         & 73.59   &   75.36  & 75.38        & 74.91         & 74.65        & 75.28                       \\
object class           & 87.41 & 85.9  & 86.33 & 77.17            & 86.16        & 84.03         & 82.74    &  87.69  & 86.69        & 87.1          & 87.36        & 87.37                       \\
multiple objects       & 65.09 & 65.38 & 65.26 & 45.85            & 66.01        & 58.78         & 58.84    &  66.36  & 67.96        & 66.17         & 66.07        & 65.11                       \\
human action           & 98.2  & 96.8  & 98 & 93               & 97.8         & 97            & 96.8     &  97.40  & 97           & 97            & 96.6         & 97.2                        \\
color                  & 86.84 & 86.12 & 86.95 & 85.63            & 87.16        & 84.72         & 85.24    &  87.82  & 88.16        & 87.07         & 85.1         & 87.66                       \\
spatial relationship   & 55.03 & 54.87 & 53.8 & 42.97            & 57.6         & 58.77         & 56.69    &  54.73  & 54.23        & 54.23         & 54.54        & 54.78                       \\
scene                  & 66.46 & 63.78 & 64.91 & 55.87            & 65.6         & 61.59         & 61.04    &  63.46  & 63.71        & 62.28         & 61.96        & 63.73                       \\
appearance style       & 81.13 & 81.87 & 80.96 & 80.39            & 81.66        & 81.83         & 81.76    &  80.85  & 81.13        & 81.41         & 81.8         & 82.01                       \\
temporal style         & 66.87 & 64.84 & 66.48  & 62.55            & 66.62        & 66.43         & 65.58    &  66.57  & 66.48        & 66.24         & 65.77        & 66.1                        \\
overall consist    & 73.46 & 72.25 & 73.46 & 70.99            & 74.18        & 73.68         & 73.63    &  73.35  & 73.08        & 72.66         & 72.66        & 73.6                       \\
\rowcolor{orange!10} \textbf{quality score}          & 82.2  & 81.39 & 82.07 & 78.98            & 81.22        & 80.43         & 79.28    &  82.19  & 82.22        & 81.91         & 81.48        & 81.48                       \\
subject consist    & 92.55 & 92.81 & 92.49 & 90.72            & 91.92        & 91.41         & 90.97    &  92.49  & 92.28        & 92.11         & 91.85        & 91.8                        \\
background consist & 94.48 & 94.19 & 94.42 & 93.97            & 94.07        & 94.06         & 94.14    &  94.34  & 94.23        & 94.04         & 93.64        & 93.87                       \\
temporal flickering    & 93.9  & 93.58 & 94.2 & 95.47            & 93.66        & 93.93         & 93.88    &  93.74  & 93.34        & 93.01         & 92.58        & 93.44                       \\
motion smoothness      & 92.8  & 93.04 & 93.07 & 93.62            & 91.9         & 91.6          & 90.87    &  92.62  & 91.94        & 91.25         & 90.39        & 91.56                       \\
dynamic degree         & 38.47 & 36.07 & 38.05 & 30               & 35.84        & 33.61         & 29.03    &  38.61  & 39.45        & 38.89         & 38.89        & 37.91                       \\
aesthetic quality      & 60.69 & 59.19 & 60.78 & 57.09            & 59.95        & 58.78         & 57.95   &   60.66  & 60.59        & 60.4          & 59.89        & 59.79                       \\
imaging quality        & 61.44 & 60.15 & 60.45 & 52.5             & 60.6         & 59.42         & 58.51   &   61.79  & 62.59        & 62.7          & 62.39        & 61.21                       \\
\bottomrule
\end{tabular}   
\label{tab:detail-cogvideo-vbench}
\end{table}

\textbf{Configure Details for Wan.} C1 in the PAB represents the default configuration, while C2 and C3 apply \texttt{block\_skip\_range=8} with \texttt{timestep\_skip\_range} set to $[100, 950]$ and $[100, 970]$, respectively. For TeaCache (denoted as TC in the table), configurations C1 and C2 correspond to \texttt{l1\_distance\_thresh values} of $0.14$ and $0.2$. For SRDiffusion (denoted as SRD in the table), configurations C1, C2, and C3 use $\delta$ values of 0.002, 0.01, and 0.03, respectively. The $SRD_{TC}$ configuration uses $\delta=0.01$ in combination with TeaCache with a threshold of $0.14$. 

\textbf{Configure Details for CogVideoX.} C1 in the PAB represents the default configuration, while C2 apply \texttt{block\_skip\_range=8} with \texttt{timestep\_skip\_range=[100,900]}. For TeaCache (denoted as TC in the table), configurations C1, C2, C3 correspond to \texttt{l1\_distance\_thresh values} of $0.1$, $0.15$ and $0.2$. For SRDiffusion (denoted as SRD in the table), configurations C1, C2, C3 and C4 use $\delta$ values of 0.008, 0.01, 0.015 and 0.03, respectively. The $SRD_{TC}$ configuration uses $\delta=0.01$ in combination with TeaCache with a threshold of $0.1$. 

\subsection{Distribution for Adaptive Switch Steps}
\label{sec:distribution-adaptive}

We analyzed the step distribution under different $\delta$ values using the standard prompt set of VBench, focusing on Wan 480p, Wan 720p, and CogVideoX 480p. Box plots were used to visualize the data. We observed that a larger $\delta$ leads to earlier switching, resulting in a higher acceleration ratio, while a smaller $\delta$ causes later switching, thereby staying more faithful to the original output of the Sketching Model. Notably, for different prompts, the switching step range of Wan is significantly wider than that of CogVideoX, which is consistent with our observations during the design of the evaluation metric, referring Figure \ref{fig:diff_values}.

\begin{figure}[hbt]
\centering
\includegraphics[width=\columnwidth]{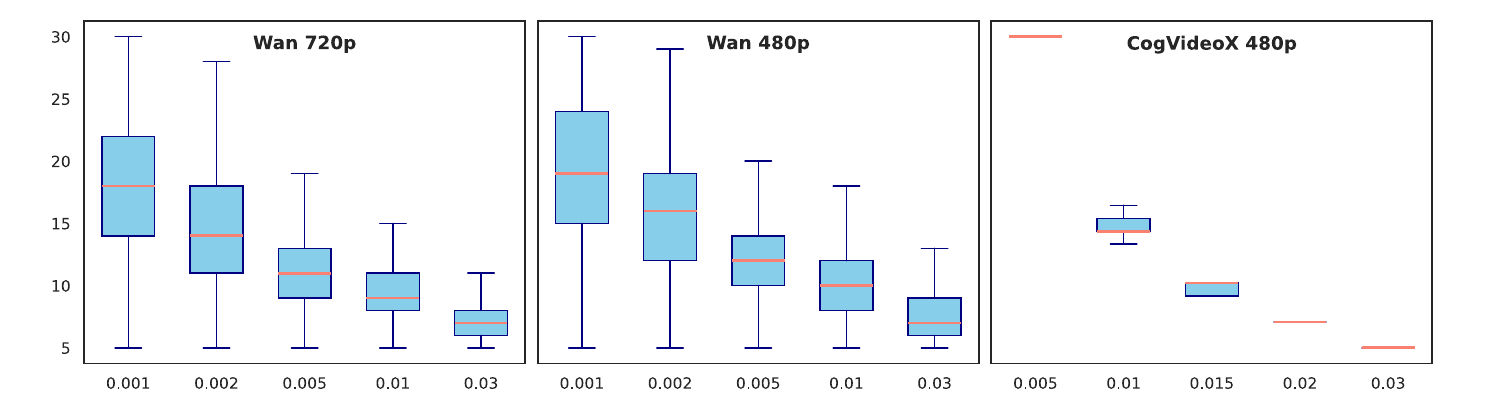}
\caption{Distribution for Adaptive Switch Steps.}
\label{fig:adapt-step-info}
\end{figure}

\subsection{More Visualization Results}
\label{sec:appendix-viz}

\begin{figure}[hbt]
\centering
\includegraphics[width=\columnwidth]{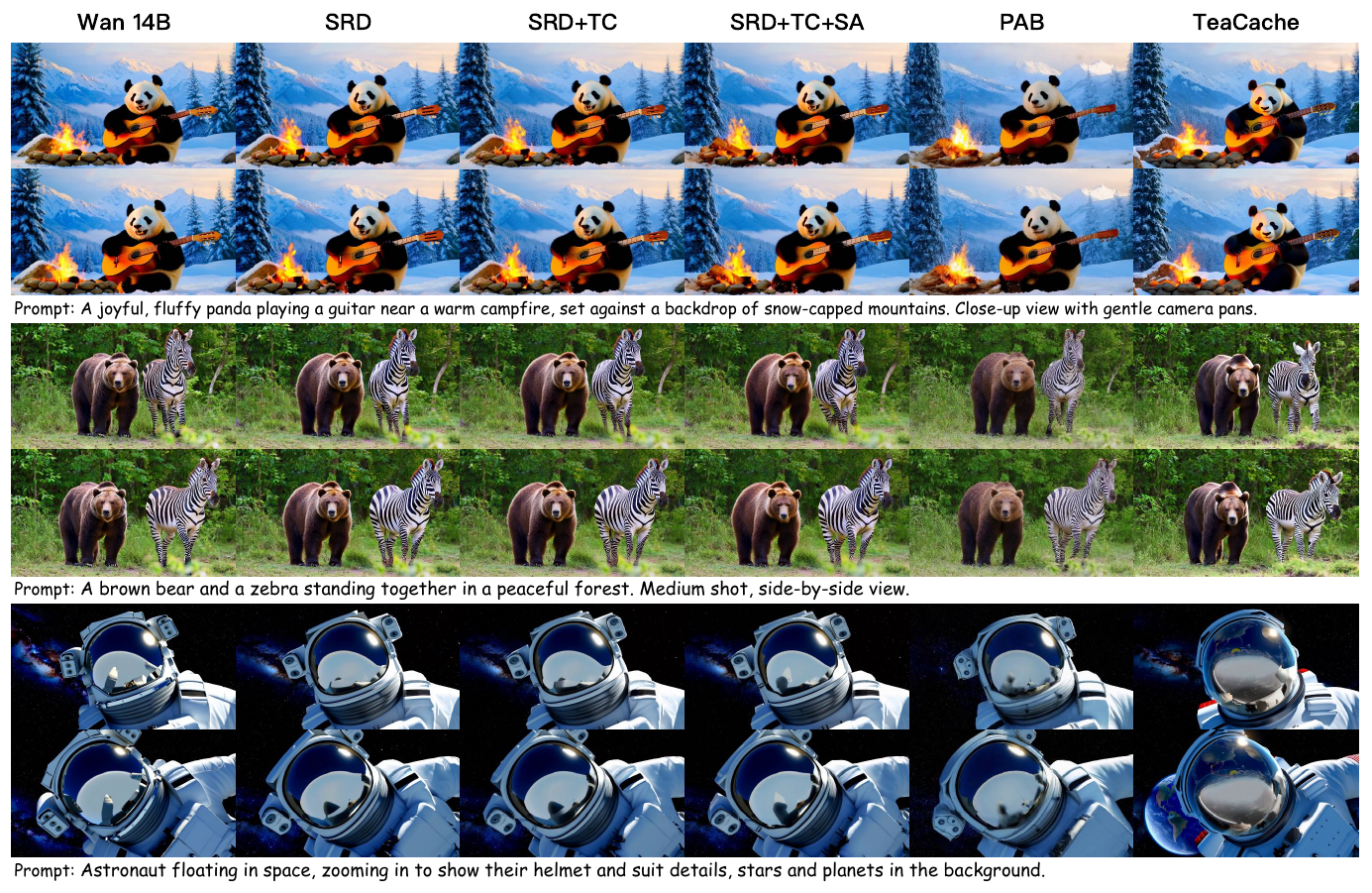}
\caption{Visualization Results of Wan model for VBench Prompts. (SDR: SRDiffusion, TC: TeaCache, SA: SageAttention)}
\label{fig:wan-vbench-viz}
\end{figure}

\begin{figure}[bht]
\centering
\includegraphics[width=\columnwidth]{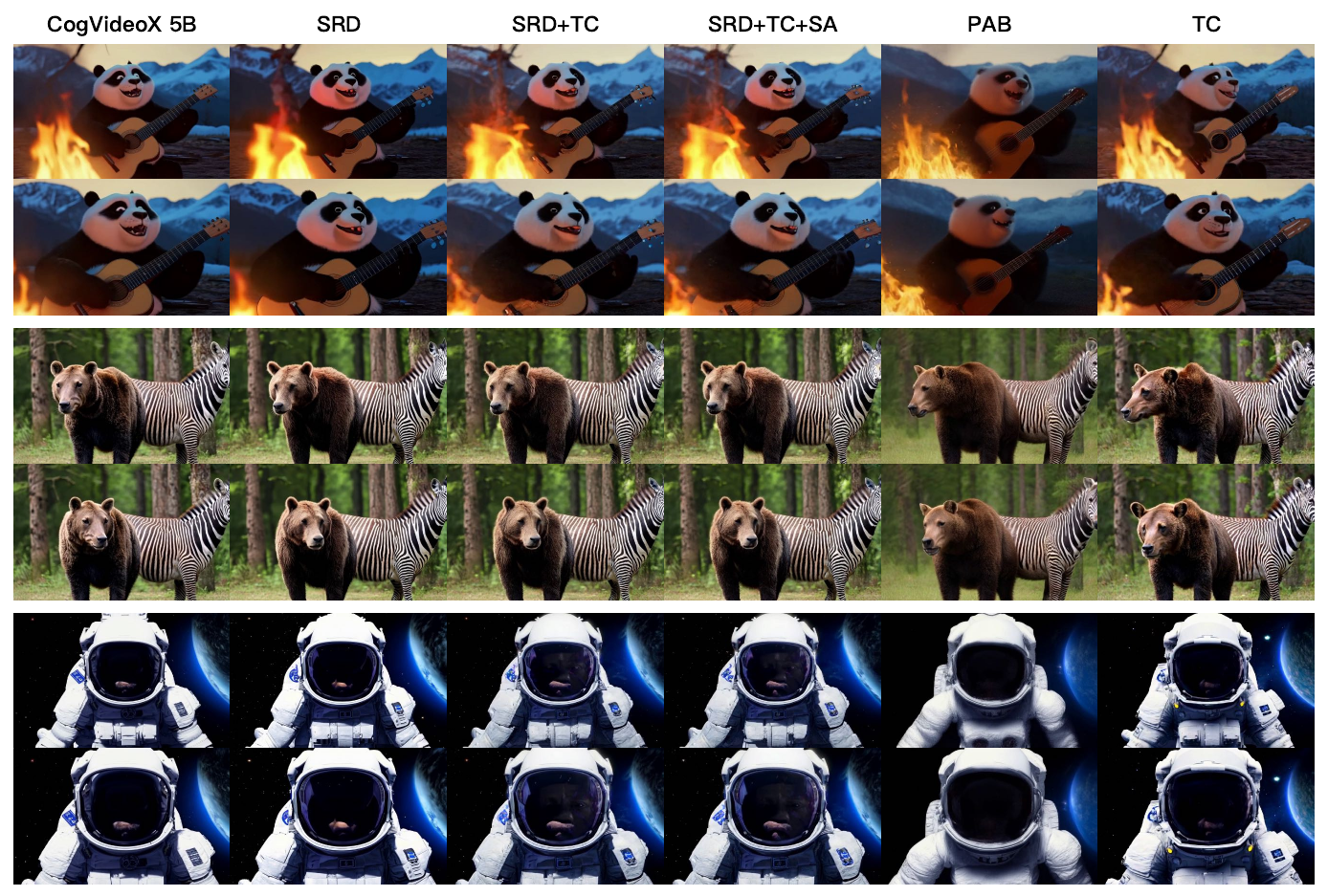}
\caption{Visualization Results of CogVideoX model for VBench Prompts, prompt same from Figure \ref{fig:wan-vbench-viz}. (SDR: SRDiffusion, TC: TeaCache, SA: SageAttention)}
\label{fig:cog-vbench-viz}
\end{figure}

In this section, we present several visual examples of video generation results:

\textbf{Figures~\ref{fig:wan-vbench-viz} and~\ref{fig:cog-vbench-viz} showcase randomly selected samples generated by Wan and CogVideoX on prompts from VBench.} For both the Wan and CogVideoX models, the configurations used for visualization include $SRD_{C2}$, $SRD_{TC}$, $SRD_{TC}$ with SageAttention, $PAB_{C2}$, and $TC_{C1}$. Detailed parameter settings can be found in Appendix~\ref{sec:detail-vbench-score}. Compared to PAB and TeaCache, SRDiffusion demonstrates a better ability to follow the original model's generation. It produces noticeably more consistent results in terms of composition and motion.

\begin{sidewaysfigure}
    \centering
    \includegraphics[width=\textwidth]{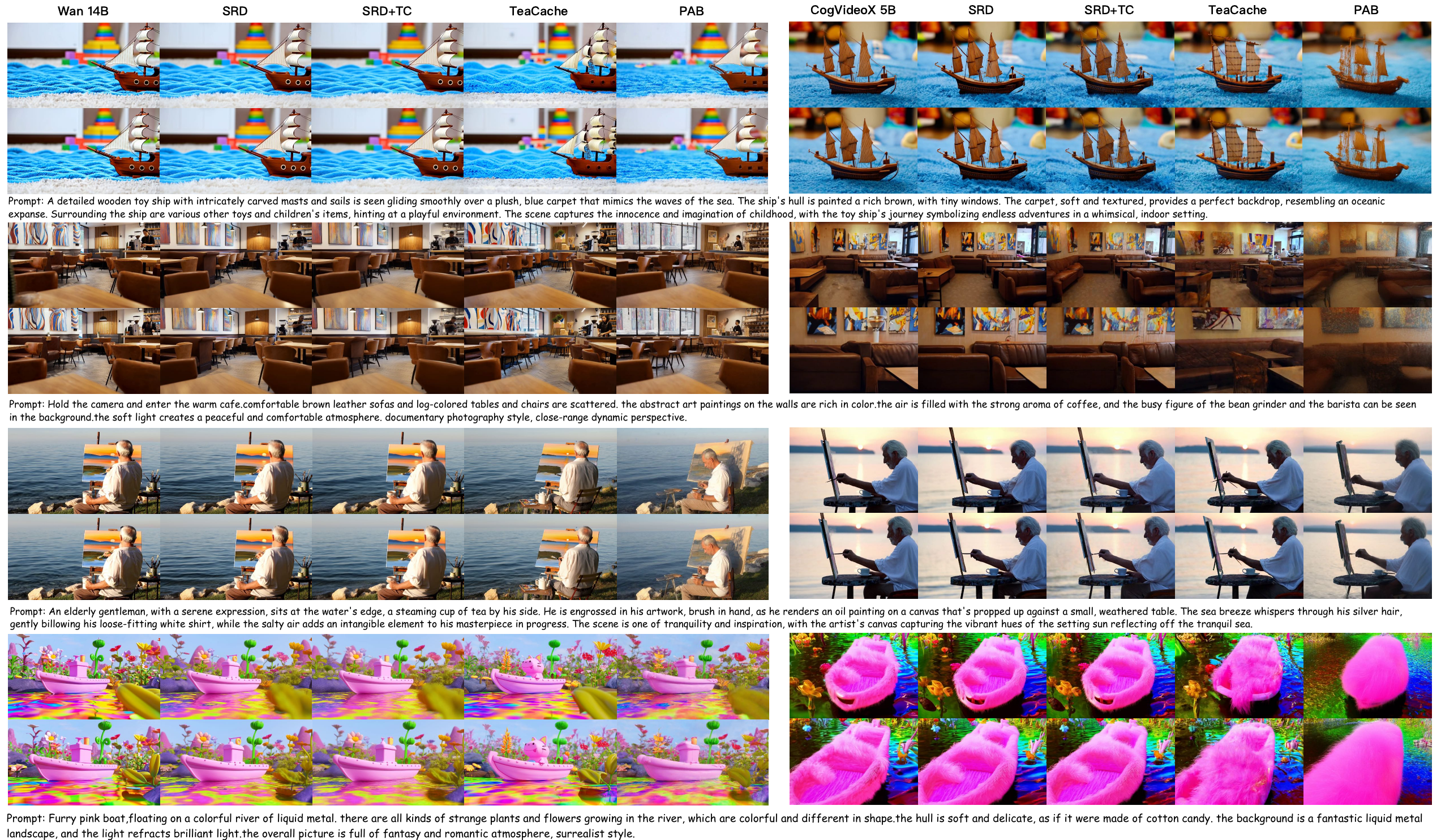}
    \caption{Visualization Results of Wan and CogVideoX model for Challenging Prompts. (SDR: SRDiffusion, TC: TeaCache, SA: SageAttention)}
    \label{fig:difficult-prompt}
\end{sidewaysfigure}

\textbf{Figure~\ref{fig:difficult-prompt} illustrates the outputs of Wan and CogVideoX on a set of challenge prompts.} The prompts in VBench are relatively simple, so we collected more challenging prompts from the open-source community for further evaluation. In this part, Wan was evaluated at 720p resolution, while CogVideoX was tested at 480p (as it only supports 480p). We observed that, compared to PAB and TeaCache, SRDiffusion better preserves the generation quality of the original model. For example, in the first prompt, both the sails and the background are more accurately rendered. In the second prompt, the structure of the house and the position of the picture frame are more similar. In the third prompt, the elderly man's appearance and the subject of his painting are better consistency. In the fourth prompt, the style of the boat and the background are more faithfully depicted.

On these more challenging prompts, SRDiffusion demonstrates a stronger ability to follow the original model’s generation while adhering more closely to the given instructions and producing finer details. This improved detail generation may stem from the fact that, unlike the baseline methods which skip certain computations, SRDiffusion employs a small rendering model that retains more reliable detail generation capabilities.

\begin{sidewaysfigure}
    \centering
    \includegraphics[width=\textwidth]{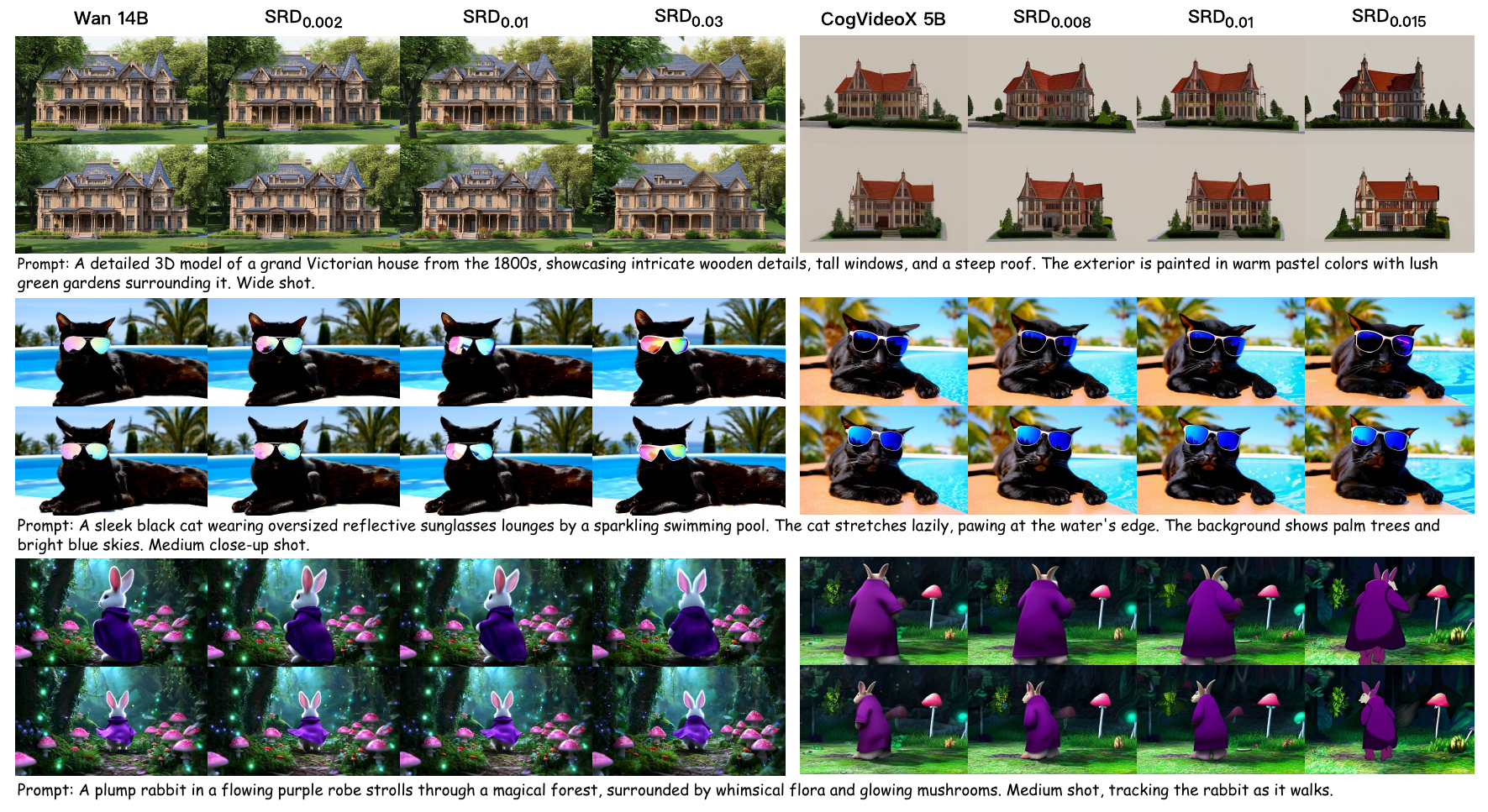}
    \caption{Comparison of visualization results from SRDiffusion using various $\delta$ values and the original model output. (SDR: SRDiffusion)}
    \label{fig:different-delta}
\end{sidewaysfigure}

\textbf{Figure~\ref{fig:different-delta} displays video generation results from SRDiffusion under different values of the parameter $\delta$.} For Wan, we visualized the results under three different settings: $\delta = 0.002$, $0.01$, and $0.03$. For CogVideoX, we tested three values as well: $\delta = 0.008$, $0.01$, and $0.015$. We observe that smaller $\delta$ values tend to more closely follow the outputs generated by the large model. For example, in the second prompt of Wan, the style of the sunglasses differs when $\delta = 0.03$. However, it's worth noting that in most cases, such differences do not indicate a decline in generation quality.

\end{document}